\documentclass{emulateapj}
\usepackage{graphicx,natbib,apjfonts}
\usepackage{amsmath}
\usepackage{amssymb}
\usepackage{epstopdf}
\usepackage{xfrac}
\newcommand{\mdotedd}{$\dot{M}_\text{Edd}$}
\newcommand{\mdot}{$\dot{M}$}
\newcommand{\XSPEC}{{\sc XSpec}}
\slugcomment{ApJ (accepted)}
\begin{document}

\date{\today}
\title{Evidence for accretion rate change during type I X-ray bursts}
\author{
Hauke Worpel\altaffilmark{1,2},
Duncan K. Galloway\altaffilmark{1,2,3},
Daniel J. Price\altaffilmark{1,2}
}
\altaffiltext{1}{Monash Centre for Astrophysics, Monash University, Clayton, Victoria 3800, Australia}
\altaffiltext{2}{School of Mathematical Sciences, Monash University, Clayton, Victoria 3800, Australia}
\altaffiltext{3}{School of Physics, Monash University, Clayton, Victoria 3800, Australia}

\begin{abstract}
The standard approach for time-resolved X-ray spectral analysis of thermonuclear bursts involves subtraction of the pre-burst emission as background. This approach implicitly assumes that the persistent flux remains constant throughout the burst. We reanalyzed 332 photospheric radius expansion bursts observed from 40 sources by the  Rossi X-ray Timing Explorer, introducing a multiplicative factor $f_a$  to the persistent emission contribution in our spectral fits. We found that for the majority of spectra the best-fit value of $f_a$ is significantly greater than 1, suggesting that the persistent emission typically increases during a burst. Elevated $f_a$ values were not found solely during the radius expansion interval of the burst, but were also measured in the cooling tail. The modified model results in a lower average value of the $\chi^2$ fit statistic, indicating superior spectral fits, but not yet to the level of formal statistical consistency for all the spectra.
 We interpret the elevated $f_a$ values as an increase of the mass accretion rate onto the neutron star during the burst, likely arising from the effects of Poynting-Robertson drag on the disk material. We measured an inverse correlation  of $f_a$ with the  persistent flux, consistent with theoretical models of the disc response. We suggest that this modified approach may provide more accurate burst spectral parameters, as well as offering a probe of the accretion disk structure. 

\end{abstract}

\maketitle

\section{Introduction}
\label{sec:Intro}
Thermonuclear (type I) X-ray bursts arise from the unstable ignition of accreted H/He near the surface of an accreting neutron star in a low-mass X-ray binary (LMXB) (e.g., \citealt{FujimotoEtAl1981, StrohmayerBildsten2006}). Gas accreted from a low-mass stellar companion accumulates on the surface of the neutron star, where it is compressed and heated hydrostatically. When the temperature and pressure are high enough a thermonuclear explosion is triggered. These events are observed as a sudden increase in X-ray luminosity to many times the persistent level (e.g., \citealt{LewinEtAl1993,StrohmayerBildsten2006}). Type I bursts have been detected from over  100 sources in our Galaxy\footnote{See http://www.sron.nl/~jeanz/bursterlist.html}. Typical bursts exhibit rise times of 1 to 10 seconds, durations of a few tens of seconds to a few minutes and have total energy outputs of $10^{39}-10^{40}$ erg.

 Analyses of type I bursts typically make a number of implicit assumptions,  namely:
\begin{itemize}
\item The total source spectrum consists of two additive components: one (the "burst component") arising from nuclear burning and the other ("persistent emission") arising from accretion.
\item The burst component has the same spectral shape for all bursts from all sources.
\item That shape is a blackbody, with variable temperature and normalization.
\item The persistent emission's spectral shape does not change during a burst and is identical to its pre-burst shape.
\item The persistent emission's intensity does not change during a burst and is identical to its pre-burst intensity.
\end{itemize}

 Making the last two of these assumptions allows the subtraction of the pre-burst emission of the neutron star from the burst spectra as background (e.g., \citealt{vanParadijsLewin1986, LewinEtAl1993, KuulkersEtAl2003,GallowayEtAl2008}, hereafter G08). This approach ( hereafter referred to as the "standard approach") implicitly assumes that the accretion rate remains constant throughout a burst, but it is not obvious that this assumption is reasonable. For instance, when the flux reaches the Eddington limit, in the so-called \emph{photospheric radius expansion} (PRE) bursts, one would na\"{\i}vely expect, assuming isotropic emission, accretion to cease entirely because the outward radiation force exceeds the gravitational force.  \cite{LambMiller1995} argue that accretion ought to be shut off if the luminosity exceeds Eddington anywhere in the accretion flow, not necessarily just at the stellar surface. The effect of radiation halting accretion flow has apparently been observed for a number of the very brightest and most vigorous bursts (e.g., \citealt{intZandWeinberg2010}), during which, for a few seconds, no X-ray flux (beyond instrumental background) is observed from the star--- though this may also be due to the atmosphere expanding so far that it obscures the emitting regions of the accretion disc. On the other hand, increased luminosity during a burst might \emph{enhance} the accretion rate, via Poynting-Robertson drag acting upon the accretion disc (\citealt{WalkerMeszaros1989,MillerLamb1996,Walker1992}, hereafter W92). At luminosities greater than $0.01L_\text{Edd}$, radiation forces have more of an effect on the accretion flow than general relativistic effects \citep{MillerLamb1993}. Since there are questions about the isotropy of the burst emission (e.g., \citealt{BoutloukosEtAl2010}), it is not obvious which of outwards pressure and radiation drag will dominate. It is therefore of importance to determine if a varying accretion rate is detectable during a PRE burst, and to quantify any variation that is detected. In this paper we attempt to measure a change in the accretion rate by performing spectral fits where the pre-burst (persistent) emission is allowed to vary during a burst,  but holding all the other assumptions of the standard approach fixed.

Some of the implicit assumptions of the standard analysis approach have been tested in previous studies: \cite{vanParadijsLewin1986} pointed out that,  if the total spectrum and persistent spectrum are both treated as blackbodies, subtracting the former from the latter will not leave a net burst spectrum that can be fit with a blackbody. This idea was followed up by \cite{KuulkersEtAl2002} in a study of the rapidly accreting GX~17+2. They did not find that accounting for this effect improved the spectral fits and concluded that the persistent emission does not originate from the same location as the burst emission on that neutron star. \cite{MunoEtAl2000} and \cite{StrohmayerBrown2002} allowed for the accretion to shut off entirely, by subtracting the instrumental background only, but did not find that the spectral fits were improved by doing so.  Recently \cite{intZandEtAl2013} studied a PRE burst from SAX~J1808.4$-$3658 using combined \emph{Chandra} and \emph{RXTE} data and found that an observed excess of photons at both low and high energies can be well described by allowing a 20-fold increase of the pre-burst persistent emission. It also may be that the persistent emission is composed of separate contributions arising from different sites. These include a boundary layer at the inner edge of the accretion disk (e.g., \citealt{KuulkersEtAl2002}), the inner regions of the accretion disk proper (e.g., \citealt{ChristianSwank1997,CackettEtAl2010}),  emission from the neutron star itself or its photosphere \citep{vanParadijsLewin1986}, and Compton scattering in an accretion disk corona \citep{WhiteHolt1982}. However, disentangling these contributions is likely to be difficult, because they are probably correlated and spectrally indistinct.

Deviations of the burst component of the spectrum from a black-body spectrum could also be present. Such spectral changes have been theoretically predicted at both the high energy (e.g., \citealt{LondonEtAl1984, LondonEtAl1986}), and low energy (e.g., \citealt{Madej1991a,MadejEtAl2004}) ends of the X-ray spectrum. However, the literature is divided as to whether these are actually present in observations. Excess photons at high energy have been reported in bursts from 4U~2129+11 \citep{vanParadijsEtAl1990} and GX~17+2 \citep{KuulkersEtAl2002}. On the other hand, pure black-bodies have been found to give generally good results up to the present time (e.g., \citealt{GuverEtAl2012}) and some authors argue that they  are consistent with black-bodies to extremely high confidence \citep{BoutloukosEtAl2010}.  One might further divide the burst emission into contributions from a continuum, and discrete spectral features superimposed upon it such as emission lines and absorption edges. Continuum changes are likely to be present throughout all stages of a burst (e.g., \citealt{SuleimanovEtAl2011}). Changes in the spectral features are thought to be largely confined to the Eddington-limited radius expansion period as these are thought to be due to ashes from nuclear burning being mixed into the expanding envelope \citep{WeinbergEtAl2006}, and have been detected in the so-called superexpansion bursts \citep{intZandWeinberg2010}. However, the non-detection of spectral features in a PRE burst from SAX~J1808.4$-$3658 by \cite{intZandEtAl2013} suggests that such features may be too weak to be detected by currently available instruments: this source is the brightest PRE burster (see Table \ref{tab:NHtable}) and the burst in question was observed by two X-ray observatories. \cite{GallowayEtAl2010} previously used \emph{Chandra} spectra to search for spectral features in PRE bursts from 4U~1728$-$34, without success. Either the sought-after features are too weak to be detected, or they are not present in every burst.

If the accretion rate varies for any bursts, we expect it to be for the PRE bursts, although there is obviously a possibility that outward radiation pressure and radiation drag will partially negate each other;  this makes them the most stringent tests of the idea that accretion rate might increase. They are the most luminous bursts for any given source and therefore should produce the largest radiation forces. They will also have the best signal-to-noise of bursts from any given source. These bursts are of interest because they can be used to probe the structure of LMXB systems. Since they reach the Eddington luminosity of the neutron star, the atmosphere is no longer bound to the surface of the star and expands. The luminosity is thought to remain within a few percent of the local Eddington luminosity throughout the radius expansion (e.g., \citealt{HanawaSugimoto1982, EbisuzakiEtAl1983, Titarchuk1994}). Thus, radius expansion bursts can in principle be used to measure the surface gravitational redshift of the neutron star and thereby its compactness (e.g., \citealt{DamenEtAl1990, OzelEtAl2009, GuverEtAl2010, SteinerEtAl2010}), giving insights into the neutron star equation of state \citep{LattimerPrakash2007}.  Radius expansion bursts can also be used to determine the distance to the bursting star \citep{BasinskaEtAl1984}, making them potentially useful as standard candles (e.g., \citealt{KuulkersEtAl2003}).

The structure of this paper is as follows. In \S \ref{sec:Data} we describe the data and its collection. In \S \ref{sec:OneBurst} we develop and implement a modified spectral analysis method, and apply it to a single burst to demonstrate its feasibility. In \S \ref{sec:ManyBursts} we apply the method to all the PRE bursts in our sample to build a statistical picture of the phenomenon. In \S \ref{sec:SpecShape} we investigate the effect of  allowing \emph{non-thermal} burst emission of fixed shape. In \S \ref{sec:WalkerComparison} we compare our results to theoretical predictions made by \cite{Walker1992}. In \S \ref{sec:Discussion} we discuss our results and place them in the context of previous work. 

\section{Data collection and reduction}
\label{sec:Data}

We used observational data from the \emph{Rossi X-Ray Timing Explorer} (\emph{RXTE}), publicly available through the High-Energy Astrophysics Science Archive Research Centre (HEASARC),\footnote{See http://heasarc.gsfc.nasa.gov} dating from shortly after the satellite's launch on December 30, 1995 to the end of the \emph{RXTE} mission on January 3, 2012. \emph{RXTE} carries three instruments for detecting X-rays. The All-Sky Monitoring camera (ASM) consists of three Scanning Shadow Cameras sensitive to photons with energies between 1.5 and 12~keV  with a field of view of approximately two degrees, and performed 90s  step-stare observations of most of the sky every 96 minutes \citep{LevineEtAl1996}. The Proportional Counter Array (PCA) consists of five PCUs sensitive to photons with energies between 2 and 60~keV and has a field of view of approximately one degree \citep{JahodaEtAl2006}.  \emph{RXTE} also carries the High Energy X-ray Timing Experiment (HEXTE), a collection of scintillation counters with a one degree field of view \citep{RothschildEtAl1998}, but we do not use data from HEXTE in this work.

We took as the basis for our sample the G08 catalog of bursts. The G08 sample identifies PRE bursts according to the criteria set out in G08 (their section 2.3). Briefly, these criteria define a PRE burst as one that i) reaches a local maximum  blackbody normalization $K_{\rm{bb}}$ at or near the moment of peak flux, ii) has declining values of $K_{\rm{bb}}$ after this time, and iii) attains its lowest blackbody temperature at the maximum $K_{\rm{bb}}$. At the time of publication of G08, the catalog contained 254 PRE bursts; since then, a further 118 PRE bursts have been observed, giving a total of 372\footnote{see \url{ burst.sci.monash.edu/minbar } }. 

Unless otherwise stated, the data analysis procedures are as in G08. For those bursts for which suitable datamodes were available, time-resolved spectra in the range 2--60~keV covering the burst duration were extracted on intervals beginning at 0.25~s during the burst rise and peak. The bin size was gradually increased into the burst tail to maintain roughly the same signal-to-noise level. The {\it RXTE}\/ Proportional Counter Units (PCUs) are subject to a short ($\approx 10\ \mu$s) interval of inactivity following the detection of each X-ray photon. This ``deadtime'' reduces the detected count rate below what is incident on the detector (by approximately 3\% for an incident rate of 400~count~s$^{-1}$~PCU$^{-1}$). We calculated for each measured spectrum an effective exposure, taking into account the fraction of events lost during deadtime, following the approach recommended by the instrument team\footnote{see \url{ http://heasarc.gsfc.nasa.gov/docs/xte/recipes/pca\_deadtime.html } }. Contributions to the dead time fraction arise from coincidence and particle events as well as source and background photons.

We re-fit the spectra over the energy range
2.5--20~keV using the revised PCA response matrices, v11.7\footnote{see \url{ http://www.universe.nasa.gov/xrays/programs/rxte/pca/doc/rmf/pcarmf-11.7 } } and adopted the recommended systematic error of 0.5\%. The fitting was undertaken using {\sc XSpec} version 12. In order to accommodate spectral bins with low count rates, we adopted Churazov weighting.

We modelled the effects of interstellar absorption, using a multiplicative model component ({\tt wabs} in {\sc XSpec}), with the column density $N_{\rm H}$ frozen. The $N_H$ values are drawn from the literature, preferentially from studies of neutron stars using instruments sensitive at lower X-ray energies than \emph{RXTE}. They are listed in Appendix A. In the original analysis carried out by G08, the neutral absorption was determined separately for each burst, from the mean value obtained for spectral fits carried out with the $N_{\rm H}$ value free to vary. This has a negligible effect on the burst flux, but can introduce spurious burst-to-burst variations in the blackbody normalization.

\section{Method}
\label{sec:OneBurst}

Our revised analysis is to fit the burst spectra with a two-component model consisting of a black-body, representing the burst emission, and a model of the pre-burst persistent emission, representing the emission due to accretion of material onto the neutron star, with a prefactor $f_a$ left free in the fits. {As we cannot distinguish contributions to the persistent emission arising from different locations in the neutron star system, we simply assume the persistent emission is indivisible and results entirely from accretion onto the neutron star.}  As PRE bursts are likely to be the most stringent tests of changes in accretion onto the star, we select these events for our analysis.

 It is also possible that the burst emission is non-Planckian, but fitting a non-black-body spectrum requires the existence of theoretical models that can describe the data. The most recent model atmospheres of bursting neutron stars are those of \cite{SuleimanovEtAl2011}, but even these have been only partially successful \citep{ZamfirEtAl2012}, and are not intended for radius expansion spectra. In the absence of models that are demonstrably better than black-bodies, and no consensus that black-body fits really are unsuitable, we keep the assumption of thermal burst emission for the majority of our analysis. We discuss this issue further in \S \ref{sec:SpecShape}. {Similarly, we do not draw a distinction between deviations from a blackbody due to continuum or features in this paper, since spectral features are either not present in every burst or are too weak to be detected.}

In order to demonstrate our approach, we select a burst from the well-studied PRE burster 4U~1636$-$536 \citep{SwankEtAl1976}. This source is an attractive candidate for several reasons. It is a prolific burster, with 75 PRE bursts recorded by \emph{RXTE}. At a distance of approximately 6~kpc (G08) it is relatively nearby. Its average peak burst flux of $(69\pm 6)\times 10^{-9}\text{ erg}\text{ cm}^{-2}\text{ s}^{-1}$ is among the brightest sources and ensures good signal-to-noise. The hydrogen column density of $0.25\times 10^{22}\text{ cm}^{-2}$ \citep{AsaiEtAl2000} is low compared to most other sources in the catalog, minimizing the absorption corrections that have to be performed on the spectra. There are no other known LMXBs in the same field of view as 4U~1636$-$536, so confusion with the persistent emission of other sources is not an issue (see \S \ref{sec:BurstSelection} for further discussion of this problem).  Finally, 4U~1636$-$536 accretes mixed H/He \citep{GallowayEtAl2006} and is therefore representative of the majority of neutron stars in our sample (G08).

The burst we chose was detected on Jun 15, 2001 (burst ID 34 in G08). The data for this burst consists of 122 spectra recorded by \emph{RXTE} with an integration time of typically 0.25s. Of these spectra, 37 are after the beginning of the burst, which we take to be the first time that the flux exceeds 25\% of the maximum burst flux. Recording continued up to 176.75s after the burst start, but the flux had declined to the pre-burst level by 24.75s. Two of \emph{RXTE}'s PCUs, numbers 3 and 4, were active for this burst. 

\subsection{Characterizing the persistent emission}
\label{sec:CharPers}

We adopted the integrated X-ray flux for a 16-second interval prior to the  start of each burst as the persistent emission. This spectrum also includes a time-dependent contribution from instrumental (non-source) background; to estimate this contribution we used the full-mission, ``bright'' source ($>40\ {\rm count\,s^{-1}}$) models released 2006 August 6  with the \texttt{pcabackest} tool. Subsequent model fits to each persistent (and burst) spectrum used the corresponding model spectrum estimated for that burst as background. We then generated a model for the persistent emission consisting of a blackbody plus a power law, both corrected for interstellar absorption. The fits were performed with {\sc XSpec} (\citealt{Arnaud1996, DormanArnaud2001}) using \texttt{wabs*(bbodyrad+powerlaw)} as the model. The hydrogen column density was kept fixed at $0.25\times 10^{22}\text{cm}^{-2}$, the value determined in \cite{AsaiEtAl2000}.

\begin{figure}
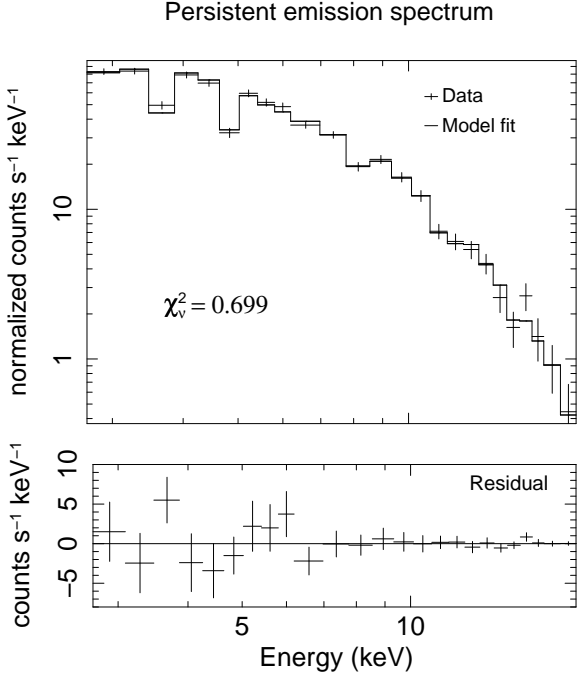

\includegraphics[angle=270, width=80mm]{figpersb.eps}\\
\includegraphics[angle=270, width=80mm]{figpersresid_b.eps}\\
\caption{Measured persistent spectrum for the Jun 15, 2001 burst from 4U~1636$-$536 and the fitted spectral model (top panel), and residuals to the data (bottom panel). The reduced $\chi^2$ is 0.699, indicating a good fit to the data.}
\label{fig1}
\end{figure}

For the persistent emission model we obtained a blackbody temperature of ($1.9^{+0.1}_{-0.1}$)~keV and a normalization of ($6.0^{+1.2}_{-1.0}$)~km$^2$/(10~kpc)$^2$. The powerlaw component had index (+3.0$^{+0.3}_{-0.2}$)~ and normalization ($2.2^{+0.8}_{-0.5}$)~keV$^{-1}$cm$^{-2}$s$^{-1}$ at 1~keV. Errors given are the $1\sigma$ confidence level. The reduced $\chi^2$ for this fit was 0.699 for 21 degrees of freedom, indicating that the model adequately describes the data. Figure \ref{fig1} shows the fit to the data and the residuals.

\subsection{Modelling the burst spectrum}
\label{sec:ModBurst}

We initially fit the burst spectra with the standard approach: a blackbody spectrum, corrected for interstellar absorption using the same hydrogen column density as above, ie. \texttt{wabs*bbodyrad} in {\sc XSpec}. We  subtracted the  measured pre-burst emission,  which includes a component that does not arise from the source (the instrumental background), and fit the resulting burst spectrum. This is the same  implementation of the standard approach as in G08. These initial fits provide standard approach fits to which we can compare ours, as well as sensible spectral parameters to seed the variable persistent emission fits.

Now we introduce the dimensionless quantity $f_a$, the persistent emission multiplicative factor, as a third variable parameter. The burst spectra are fitted again, replacing the  recorded pre-burst emission subtraction with just the instrumental background, with the model 
\begin{equation}
S(E)=A(E)\times B(E;T_\text{bb},K_\text{bb}) + f_a\times P(E)-b(E)_\text{inst},
\label{eqn:modeldescription}
\end{equation}
 where $S(E)$ is the fitted spectrum as a function of energy $E$, $A$ is the absorption correction, $B$ is the blackbody function with temperature $T_\text{bb}$ and normalization $K_\text{bb}$, $P$ is the persistent model described in \S \ref{sec:CharPers}, and $b_\text{inst}$ is the instrumental background. Note that the persistent model already  contains an absorption factor. The parameter $f_a$ is allowed to vary between -100 and 100. This allows us to track changes in the accretion rate through the factor $f_a$. Because of the 0.25~s integration time of the burst spectra, our measured peak $f_a$ values may be slight underestimates.

\begin{figure*}
\includegraphics[angle=270,width=170mm]{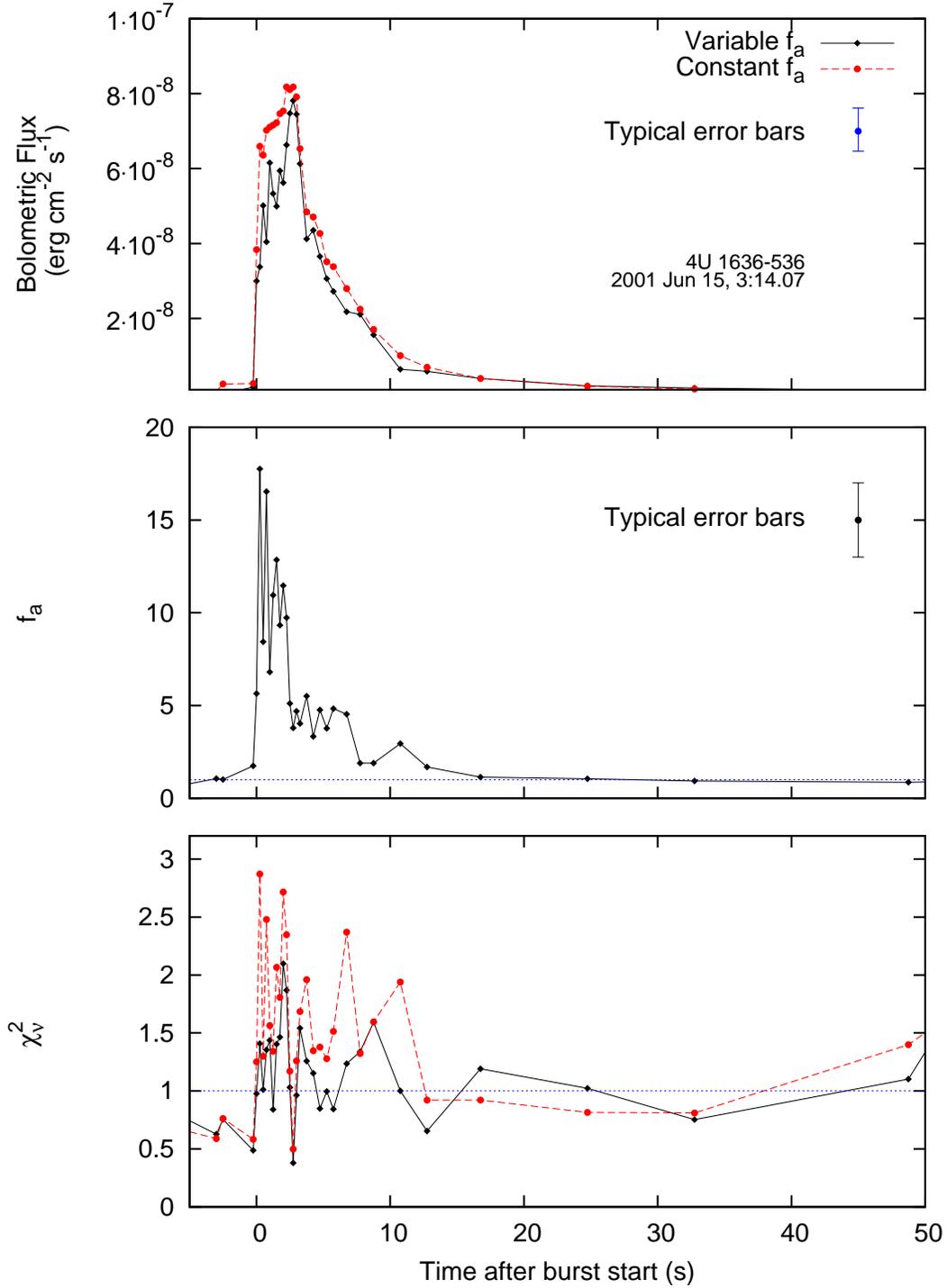}
\caption{Comparison of fitting a variable persistent emission factor, $f_a$, to the standard  approach fits for a burst from 4U~1636$-$536, where $f_a$ is a scaling factor of the pre-burst emission. The variable fit yields consistently lower burst fluxes (top panel),  though these are within the error bars. The contribution to the flux from the variable persistent emission increases to many times the pre-burst level (middle panel) during the rise of the burst.  The $f_a$ factor remains elevated for a considerable time after the total flux begins to decrease. Allowing the persistent emission to vary improves the reduced $\chi^2$ (lower panel). See Figure \ref{fighr} for the changes in normalization and temperature for this burst.}
\label{figlayered}
\end{figure*}

\begin{figure}
\includegraphics[angle=270,width=80mm]{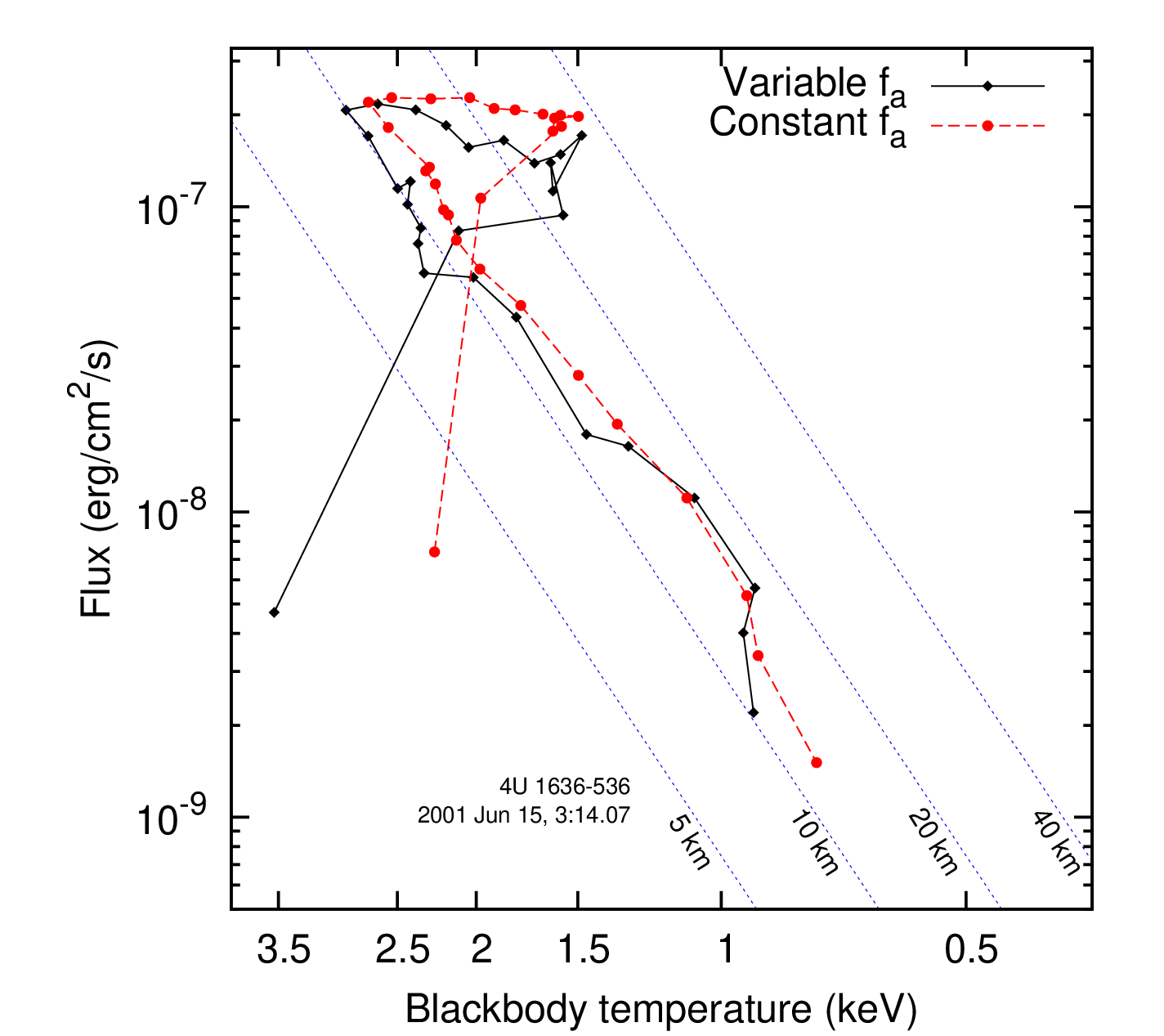}
\caption{ Flux-temperature diagram for the 2001 Jun 15 burst from 4U~1636$-$536, as measured by both the standard fit and the variable $f_a$ fit. The fluxes shown are bolometric fluxes calculated from the blackbody parameters, assuming a distance to the source of 6~kpc (G08). The variable fit is consistently slightly lower in flux,  though within the uncertainties. Also plotted are lines of constant radius for 5~km, 10~km, 20~km, and 40~km. These radii do not take into account spectral corrections (e.g., \citealt{SuleimanovEtAl2011}). The radius of the photosphere at maximum expansion is the same using both methods, but the variable $f_a$ fit gives a smaller, hotter photosphere at the touchdown point and beginning of the cooling tail.  Typical error bars are given for points near peak flux; at lower flux they are much larger.}
\label{fighr}
\end{figure}

Figure \ref{figlayered} shows the time evolution of burst bolometric flux (top panel), the $f_a$ factor (middle panel), and the reduced $\chi^2$ for both treatments (lower panel) for our selected burst. We found that the $f_a$ increases to many times the pre-burst level, peaking at $17.8\pm 4.7$ at 0.25 seconds after the burst start. The errors are $1\sigma$ significance limits determined by {\sc XSpec}. The burst bolometric flux is calculated from the blackbody parameters 
\begin{equation}
\begin{array}{rcl}
F&=&\sigma T^4_{\text{bb}}(R/d)^2\\
&=&1.076\times 10^{-11}\left(\dfrac{kT_{\text{bb}}}{1\text{ keV}}\right)^4K_{\text{bb}}\text{ erg cm}^{-2}\text{ s}^{-1},
\end{array}
\end{equation}
where $T_\text{bb}$ is the blackbody temperature, $R$ is the effective radius of the emitting surface, $d$ is the distance to the source, and $K_\text{bb}$ is the normalization of the blackbody assuming isotropic emission. The calculated flux does not include the contribution to the total flux due to the scaled persistent emission. The  burst component flux for the variable $f_a$ fits is therefore consistently lower than the  standard approach fits for the peak of the burst, where $f_a>1$. We also find that, after the beginning of the burst, the $\chi^2_\nu$ for the variable $f_a$ fits is consistently lower than for the  standard approach fit, with means of $1.26\pm0.59$ and $1.50\pm0.60$ respectively.  A Kolmogorov-Smirnov test shows that the two sets of $\chi^2_\nu$ values have a 4.0\% probability of being consistent with one another. Figure \ref{fighr} shows the flux-temperature curve for this burst.

The pre-burst persistent emission is very much fainter than the peak of the burst emission, by a factor of 35. It is possible that counting statistics in the burst spectrum could induce a spurious response in $f_a$. To investigate this possibility we examined the spectrum with the highest $f_a$, which was $17.8\pm 4.7$ recorded 0.25s after the burst start. We took the blackbody parameters for that spectrum, $kT=1.572$~keV and normalization 1034~$\text{km}^2\,/(\text{10 kpc})^{2}$, and generated $10^3$ simulated spectra consisting of an absorbed blackbody with those parameters plus the persistent model times unity. The simulated spectra also incorporated counting statistics typical of the detector. These represent a burst spectrum for which $f_a$ does not change and for which any measured deviation from $f_a=1$ must be due to noise or the fitting procedure.

\begin{figure}
\includegraphics[angle=270,width=80mm]{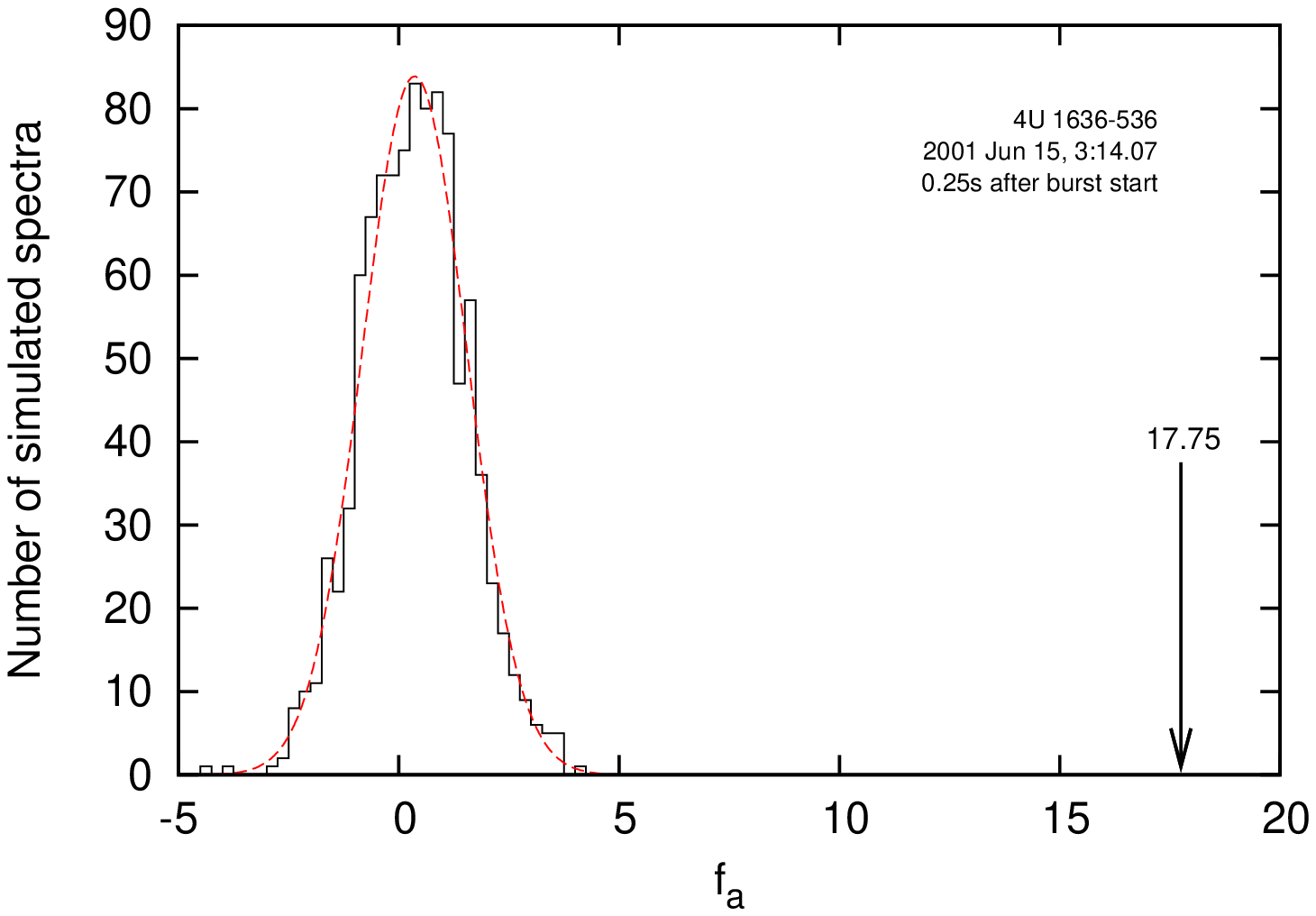}
\caption{Histogram of the measured persistent emission factor $f_a$ for 1000 simulated spectra for the burst from 4U~1636$-$536. These spectra consist of a blackbody with parameters taken from the real spectrum with the highest measured $f_a$, plus the persistent model times unity. The scatter arises from the (simulated) counting statistics. The observed spectrum is 14.6 standard deviations from the predicted $f_a$ assuming the persistent emission does not change.}
\label{figsignificance}
\end{figure}

We fit each of the simulated spectra with our variable $f_a$ model. The mean parameters for these were $kT=1.569\pm0.011$~keV, $K_\text{bb}=1061\pm 49$ km$^2/(10$kpc$)^2$ and $f_a=0.36\pm 1.18$. The mean parameters are therefore consistent with those that seeded the simulated fits. Adopting the standard deviation of the simulation values as the error on the fitted $f_a$ we found our measured $f_a$ for the real burst spectrum was 14.6 standard deviations higher than the simulated mean. We therefore conclude that the high $f_a$ values do not arise from counting statistics alone. Figure \ref{figsignificance} shows a histogram of the  measured $f_a$ for the simulated spectra compared with the value measured from the real burst spectrum.

Thus, for one burst we have found evidence for a spectral effect which can be described by varying the level of the persistent emission. This change is statistically significant for at least some of the spectra; 20 of the 37 of the spectra after the beginning of the burst had $f_a$ above unity with more than 4$\sigma$ significance  (using the above procedure, see also figure \ref{figsignificance}). The associated improvement in $\chi^2$ is of greater than $3\sigma$ significance using f-tests. We apply the new method to all the PRE bursts in our sample in the next section.

\section{Statistics of many bursts}
\label{sec:ManyBursts}

\subsection{Burst selection}
\label{sec:BurstSelection}

We restrict our sample of PRE bursts to exclude events that are unsuitable for analysis.  Three bursts were excluded for which the Standard-2 data was missing, preventing estimation of the instrumental background via \texttt{pcabackest}. 

Some burst sources, such as 4U~1728-34, AQL~X-1, and GRS~1747$-$312, lie in crowded fields containing other LMXBs. If the other source(s) were active at the time of observation, then their persistent emission could be confused with the source under observation and it would not be possible to scale only the persistent flux from the burst source. These sources need to be excluded from consideration.

We used All-Sky Monitor (ASM) data to quantify this effect\footnote{see \url{ xte.mit.edu } }. For each burst we found every source within the PCA's field of view and took the ASM counts at the time record nearest the burst start time. Since the PCA's response drops off approximately linearly with distance from the field of view's center, we multiplied the ASM counts by $1-s$, where $s$ is the angular distance of the source from the center of the field of view, in degrees. We compared all the other sources to the burst source, and if the total contribution is more than 10\% of the pre-burst flux of the burst source we excluded this burst from consideration. We found source confusion in 36 bursts. The majority of these are from 4U~1728-34, which lies very near ($\Delta\theta=0.56^\circ$) in the sky to the Rapid Burster, and which was frequently observed to burst during targeted observations of the Rapid Burster \citep{FoxEtAl2001}. 

\subsection{Modelling the persistent emission}
\label{sec:Pers-Models}

In order to measure a change in the level of persistent emission for every burst spectrum we need to construct a model for the persistent emission for each burst that can then be incorporated into the burst fits. It is not possible to simply use the detected photon counts for each energy bin because these include instrumental background.

\begin{figure}
\includegraphics[width=80mm]{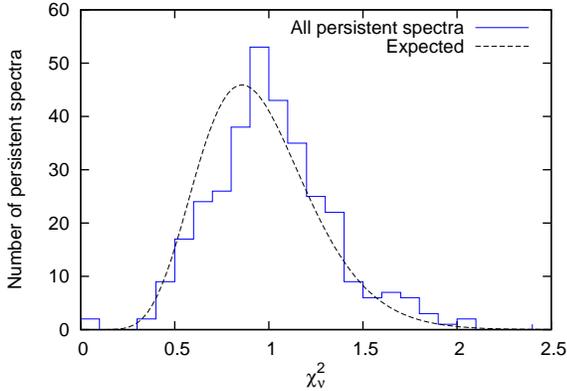}
\caption{Distribution of $\chi^2_\nu$ for the fits to the persistent emission spectra for the 332 PRE bursts (blue stepped line) and the theoretical distribution (dashed curve) for $\nu=20$. A Kolmogorov-Smirnov test showed a 8.8\% probability that the recorded distribution is consistent with the expected assuming a good fit, indicating that our model fits of pre-burst spectra are adequate.}
\label{figchi2}
\end{figure}

We found that the blackbody-plus-powerlaw model used in \S \ref{sec:OneBurst} does not give adequate fits for every burst, so we considered a set of alternative spectral models. We fit each persistent spectrum with  a range of different models. These are summarized in Table \ref{tab:Persistent-models}, and retained the fit that gave the best (ie. lowest) $\chi^2_\nu$.  We found that the six models listed in Table \ref{tab:Persistent-models} are sufficient to describe the persistent emission of all but one burst (see Figure \ref{figchi2}). We are not overly concerned with theoretical interpretations of these models-- the main objective is to get a function that matches the data-- but the literature does provide some commonly-used functions. \cite{White1986} points out the importance of scattering, mentioning that this can be modelled with a Compton spectrum or a blackbody with an added power law, and we include both in our selection of alternatives. The inclusion of a Gaussian at 6.4~keV was motivated by the detection of fluorescent Fe emission for some sources at this energy (G08). The combination of thermal bremsstrahlung with a Gaussian was found by  experimenting in \XSPEC\ with persistent spectra that could not be properly fit with any of the other five models. For models containing a Gaussian, the lower limit on the line width is set to 0.1~keV to avoid the Gaussian simply approximating a delta function that removes the error on a single energy bin. In models where the hydrogen column densities are held constant, these are given in Appendix A. The $N_H$ values  are mostly taken from the literature and references are given in Appendix A.  In persistent emission models where the hydrogen column densities are allowed to vary, the quantity $A(E)$ (see equation \ref{eqn:modeldescription}) is to be thought of as as the product of the true interstellar absorption and a multiplicative factor that corrects the shape of the persistent emission model. As $A(E)$ is a multiplicative factor it is unaffected by changes in $f_a$ and so this treatment does not introduce any systematic effects. One persistent emission spectrum,  preceding a burst from Cyg~X-2 observed at 14:29 March 27, 1996, (burst ID 2 in G08) could not be fit adequately by any of the six models, so this burst was excluded from the analysis.

The reduced $\chi^2$ for the spectral fits to the remaining 332 persistent spectra had a mean of 1.03 and a standard deviation of 0.35, with an average of $\nu=21$ degrees of freedom. The skewness was 1.00 and the kurtosis was 2.8. This distribution is therefore somewhat more skewed and significantly more peaked compared to expected theoretical values of $\sqrt{8/\nu}=0.62$ and $12/\nu=0.57$ for the skewness and kurtosis respectively. However, a Kolmogorov-Smirnov test gave a value of $D=0.069$ and a 8.8\% probability that the  measured distribution is consistent with the expected assuming a good fit. We therefore consider these persistent emission models adequate for use in the subsequent generation of burst fits. The distribution of $\chi^2_\nu$ for the fitted persistent emission spectra is shown in Figure \ref{figchi2}. 

\subsection{Fitting burst spectra}
\label{sec:Fit-burst-spectra}

As in \S \ref{sec:OneBurst}, initially we fit the burst spectra via the standard approach using an absorbed blackbody,  by subtracting the recorded pre-burst emission, ie. \texttt{wabs*bbodyrad} in {\sc XSpec}. This is the same  implementation of the standard approach as the one outlined in G08.

We then replace the  detected pre-burst emission with just the instrumental background and fit the same spectra with a blackbody corrected by the adopted interstellar extinction value plus a multiple of the persistent emission model for that burst. We fix the hydrogen column density to the values given in Appendix A. The multiple of the persistent emission $f_a$ is allowed to vary as a free parameter, along with the temperature and normalization of the blackbody, as we did in \S \ref{sec:OneBurst}.

The Levenberg-Marquardt algorithm used by {\sc XSpec} can become trapped in local $\chi^2$ minima, returning obviously unphysical results such as extremely high temperatures or normalizations. This problem can be mitigated if sensible initial values for the parameters are given rather than the {\sc XSpec} defaults, which are not appropriate in every case. We use the blackbody temperature and normalization of the blackbody obtained from the standard model, and $f_a~=~1$, to seed these fits.

Generating and fitting to 1,000 simulated spectra, as was done in \S \ref{sec:ModBurst}, for each of the tens of thousands of burst spectra would be computationally prohibitive. To avoid needless work, we excluded spectra for which the variable $f_a$ fit was spurious. Only spectra with blackbody temperatures between 0.1 and 5.0 keV, normalizations of less than $10^6\ {\rm km}^2/(10\text{km})^2$, and bolometric fluxes less than $10^{-6}\ {\rm erg\,cm^{-2}\,s^{-1}}$ were retained. We also excluded $f_a$ determinations for which {\sc XSpec} encountered fitting errors such as non-monotinicity or reaching the parameter limits. This left 26,113 burst spectra out of the original 41,282.  Almost all (>99\%) of the discarded spectra were recorded before the beginning of the burst or very late in the cooling tail (i.e. stages 0 or 4 in Table \ref{tab:stage-defns}; see also \S \ref{sec:burststages}), and the failure to obtain spectral fits can be attributed to very low photon counts for these spectra. We also reduced the number of simulated spectra for each measured spectrum from 1,000 to 320.

 As a further check that $f_a$ is measuring a real spectral effect, we took the highest $f_a$ spectrum for each burst and performed an f-test on it, comparing the $\chi^2$ statistic from the standard approach and variable $f_a$ fits. Because $f_a\times P(E)$ is an additive component (see equation \ref{eqn:modeldescription}), f-tests are a suitable test (e.g., \citealt{OrlandiniEtAl2012}). We found that, of these, 165 had detections of greater than $3\sigma$ significance and 65 had detections of better than $4\sigma$ significance.

We define the \emph{burst stage}: the \emph{pre-burst stage} consists of all times
before the beginning of the burst, the \emph{expansion stage} occurs from
the beginning of the burst to the time of maximum normalization, the
\emph{contraction stage} is from the time of the maximum normalization up to and
including the touchdown time, and the \emph{cooling tail stage} is all
times after the touchdown time but before the bolometric flux drops back below one quarter of the maximum flux. We neglect spectra after this time. We refer to the expansion and contraction stages collectively as the \emph{Eddington-limited stage}. These stages are summarized in Table \ref{tab:stage-defns}.

\begin{deluxetable*}{lcl}
\tablecolumns{2}
\tablecaption{XSPEC models for fitting the persistent emission}
\tablehead{
  \colhead{ XSPEC model } &
  \colhead{Number} &
  \colhead{ Notes } 
  \\
  & 
  \colhead{of spectra} &
}  
\startdata               
\texttt{ wabs*(bbodyrad+powerlaw) } & 127 & nH fixed to values in Appendix A \\
\texttt{ wabs*(bbodyrad+powerlaw+gauss) } & 62 & nH fixed, Gaussian energy set to 6.4keV\\
\texttt{ wabs*(compTT)\tablenotemark{a} } & 37 & nH fixed \\
\texttt{ wabs*(bbodyrad+powerlaw) } & 18 & nH allowed to vary \\
\texttt{ wabs*(gauss+bremss) } & 31 &all parameters variable\tablenotemark{b} \\
\texttt{ wabs*(bbodyrad+diskbb\tablenotemark{c}) } & 57 & all parameters variable\tablenotemark{b} \\
\hline
Total usable spectra& 332 & \\
\hline
Rejected due to source confusion & 36 & Other active sources in field\\
No background data & 3 & \\
No good persistent model fit & 1 & Minimum $\chi^2_\nu > 5$
\enddata
\tablenotetext{a}{See \cite{Titarchuk1994b}}
\tablenotetext{b}{All parameters are variable for the generation of persistent emission models. Their values are subsequently frozen for the burst spectral fits in \S \ref{sec:Fit-burst-spectra}}
\tablenotetext{c}{See \XSPEC\ manual (\url{http://heasarc.gsfc.nasa.gov/xanadu/xspec/manual/manual.html}) and references therein}
\label{tab:Persistent-models}
\end{deluxetable*}

\subsection{$f_a$ as a function of burst stage}
\label{sec:burststages}

\begin{deluxetable*}{cll}

\tablecolumns{2}
\tablecaption{Stages of a PRE burst}
\tablehead{
  \colhead{ Stage number } &
  \colhead{ Stage name} &
  \colhead{ Description } 
}  
\startdata               
0 & Pre-burst & Flux has not yet reached 1/4 of the maximum \\
1 & Expansion & Normalization has not yet reached maximum \\
2 & Contraction & Radius has reached maximum, $kT$ has not yet reached maximum\\
3 & Cooling tail & Flux still above 1/4 maximum\\
4 & Post-burst & Flux has dropped below 1/4 maximum\\
\enddata
\label{tab:stage-defns}
\end{deluxetable*}

\begin{figure}
\includegraphics[angle=270,width=80mm]{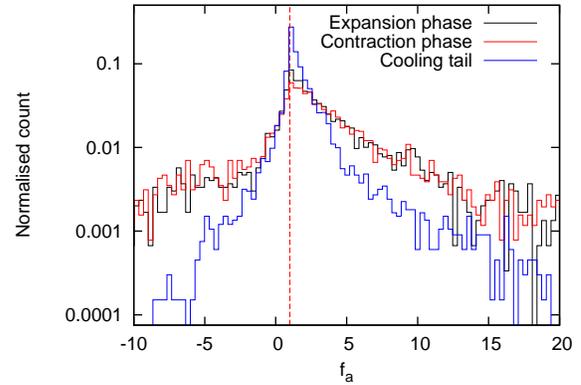}
\caption{Distribution of measured pre-burst persistent emission factor $f_a$ for cooling tail spectra and Eddington limited spectra for all PRE bursts.  The cooling tail spectra are strongly peaked around 1, while the Eddington-limited spectra are not as strongly peaked. The distributions for the Eddington limited spectra have more prominent tails, indicating a large number of elevated $f_a$ values, but there a significant number of high $f_a$ values in the cooling tail also.  The distribution for the expansion and cooling phases is consistent, suggesting that the fallback of the atmosphere does not contribute significantly to the emission.}
\label{figfahisto}
\end{figure}

Figure \ref{figfahisto} shows the distribution of persistent flux factor $f_a$ for every burst spectrum (ie. spectra from the Eddington-limited and cooling tail stages). The distribution of $f_a$ has a wider spread for the Eddington-limited spectra,  so that there is a larger fraction of high $f_a$ spectra during radius expansion than for the cooling tail. However, the population of high $f_a$ spectra in the cooling tail is not negligible. This suggests that elevated $f_a$ is not only a result of the burst being Eddington-limited. 

There are 451 spectra with $f_a$ less than zero, for which the $f_a$ measurement is significant to more than $3\sigma$ (via the procedure detailed in Figure \ref{figsignificance}), and which occur in the Eddington-limited phase, out of 26,113. Of these, 333 spectra come from just six bursts. These six events are summarized in Table \ref{tab:negfa}. By inspection of the flux-temperature curves we identified that four of the six are examples of \emph{superexpansion} bursts \citep{intZandWeinberg2010}, particularly powerful PRE events for which the atmosphere is expanded to much larger radii than usual. In these events the temperature of the photosphere drops out of the detection band of \emph{RXTE}, resulting in zero observed flux from the star. The other two are bursts with highly unusual flux-temperature curves and appear to consist of two consecutive expansions separated by a few seconds, followed by a cooling tail with larger blackbody radius than the maximum radius reached during expansion.

\begin{deluxetable*}{llcl}

\tablecolumns{2}
\tablecaption{Bursts with many negative $f_a$ burst spectra}
\tablehead{
  \colhead{ Source } &
  \colhead{ Date and time} &
  \colhead{ G08 burst ID } &
  \colhead{ Description }
}  
\startdata               
4U 1722$-$30  & 1996, Nov 8, 07:00 & 1 & Super-expansion burst\tablenotemark{a}\\
4U 2129+12 & 2000, Sep 22, 13:47 & 1 & Super-expansion burst\\
2S 0918$-$549 & 2008, Feb 8, 03:02 & 5\tablenotemark{b} & Super-expansion burst\tablenotemark{c}\\
4U 1722$-$30  & 2008, Mar 1, 16:18& 4\tablenotemark{b} & Super-expansion burst\tablenotemark{a}\\
\hline
XB 1832$-$330 & 1998, Nov 27, 05:45 & 1 & Atypical burst profile\\
HETE J1900.1$-$2455 & 2010, Sep 20, 05:29 & 7\tablenotemark{b} & Atypical burst profile\\
\enddata
\tablenotetext{a}{This burst was also studied by \cite{intZandWeinberg2010}}
\tablenotetext{b}{This burst postdates G08; a consistent burst ID numbering scheme is assumed}
\tablenotetext{c}{This burst was also studied by \cite{intZandEtAl2011}}
\label{tab:negfa}
\end{deluxetable*}

\subsection{Distribution of $f_a$}
\label{sec:fallback}

\begin{figure}
\includegraphics[angle=270,width=80mm]{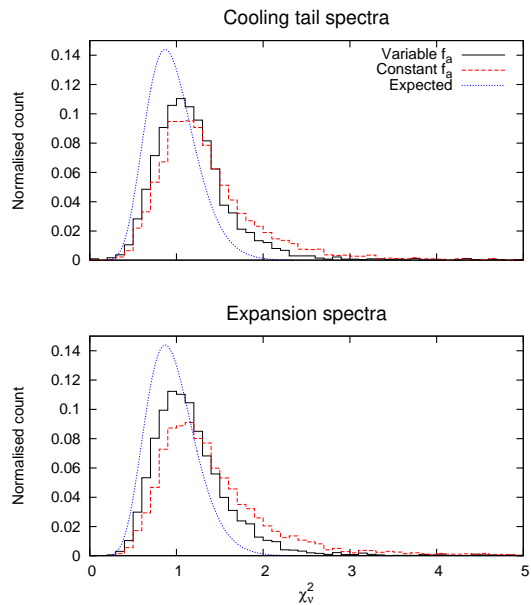}
\caption{Distribution of reduced $\chi^2$ for the variable $f_a$ vs.  standard approach spectral fits, for both cooling tail and Eddington-limited spectra. Also plotted is the theoretical distribution for 23 degrees of freedom, which is the mean for the  standard approach fits (blue line). Allowing the persistent emission to vary significantly improves the fits, but there are still significant deviations from the expected distribution for a perfect model.  The variable $f_a$ and standard approaches are not consistent with being drawn from the same distribution using a Kolmogorov-Smirnov test.}
\label{figburstchi2}
\end{figure}

We compared the distributions of $\chi^2_\nu$ to the expected distribution assuming a correct model using the Kolmogorov-Smirnov test. For the cooling tail we obtained $D=0.304$ for the  standard approach fit and $D=0.193$ for the variable fit. For the Eddington-limited spectra we obtained $D=0.331$ and $D=0.171$ respectively.  These results indicate that allowing $f_a$ to vary consistently improves the goodness-of-fit, and the improvement is more pronounced for the Eddington-limited spectra than for the cooling tail. These distributions are shown in Figure \ref{figburstchi2}. However, none of the Kolmogorov-Smirnov tests were consistent with the null hypothesis; even our variable $f_a$ fits do not adequately fit the data, though they are an improvement on the old method. Some other systematic error must yet contribute to the discrepant spectral fits, perhaps deviations in the intrinsic burst spectrum from the assumed blackbody shape.  Kolmogorov-Smirnov tests on the variable $f_a$ method against the standard approach likewise indicates that the two methods are not consistent with each other, for either the cooling tail or Eddington-limited spectra. 

\begin{figure}
\includegraphics[angle=270, width=80mm]{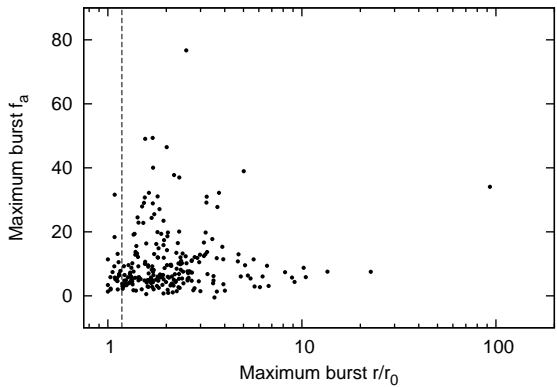}
\caption{Maximum $f_a$ plotted against reduced radius (photospheric radius divided by radius at touchdown) for each burst. The two maxima do not necessarily occur at the same time.  The dashed line shows the location of an accretion disc boundary layer at $R=1.18R_*$ \citep{PophamSunyaev2001}. The majority of photospheric expansion events exceed this radius. There is no obvious relationship between maximum $f_a$ and maximum expansion, nor is there a decrease in $f_a$ outside the accretion disc radius, suggesting that obscuration of the disc by the expanding atmosphere does not significantly affect the observed accretion emission.}
\label{figfavra}
\end{figure}

We also investigated whether the degree of radius expansion correlates with the increase in persistent emission. We define the reduced radius for an individual burst as the photospheric radius divided by the radius at the touchdown point, which corrects for the distance of the source,  assuming the isotropy factor does not change. We compared the maximum reduced radius with the maximum $f_a$ for each burst; see Figure \ref{figfavra}. These two maxima do not necessarily occur at the same moment. We found only a slight relationship between the two quantities; a Kendall $\tau$ rank correlation test found $\tau=0.138$ with $3.6\sigma$ significance. This suggests that obscuration of the inner parts of the disc does not greatly affect the observed persistent emission.

One  possible contribution to the elevated persistent flux during contraction is the fallback of the extended atmosphere. During the expansion phase, the atmosphere can be driven off at about $10^{18}$~g~s$^{-1}$ \citep{WeinbergEtAl2006}. This is, naturally, around the order of \mdotedd, the Eddington accretion rate. As an order-of-magnitude estimate we make the assumption that half the expanded mass is driven off permanently. In a burst that is Eddington-limited for 5 seconds,  the mass that is not expelled is about $2.5\times 10^{18}$~g. There are approximately 40\% more contraction spectra than expansion spectra. All the spectra at these high fluxes are taken at 0.25s intervals, so the atmosphere must generally contract more slowly than it expanded, by a factor of about 40\%. It follows that the accretion rate due to the atmosphere falling back could be as much as \mdotedd $/3$. There are bursters for which the pre-burst accretion rate is only 1\% of \mdotedd, so for these sources there should be a noticeable excess of contraction stage spectra with $f_a$ around 20-30. We have not detected any such excess (see Figure \ref{figfahisto}), so we conclude that atmosphere fallback does not contribute significantly to the accretion luminosity.  A careful comparison with non-PRE bursts would clarify this issue further, as by definition atmosphere fallback cannot occur in these events, and this will be investigated in a subsequent paper.

\section{Spectral shape changes}
\label{sec:SpecShape}
Deviations of the burst component (that is, everything except persistent emission) from a blackbody could also be reflected by a changing $f_a$, especially if these deviations manifest as an excess of photons at high energy. Such spectral changes have been theoretically predicted at both the high energy (e.g., \citealt{LondonEtAl1984, LondonEtAl1986}), and low energy (e.g., \citealt{Madej1991a,MadejEtAl2004}) ends of the X-ray spectrum. Deviations from a pure black-body are frequently inferred  during radius-expansion bursts (see, e.g., \citealt{KuulkersEtAl2002, KuulkersEtAl2003}). The influence of these deviations on $f_a$ measurements must be investigated, but this is difficult because the nature of the deviations is not known. Here we attempt to test the alternative hypothesis that the poor $\chi^2_\nu$ values  given by the standard approach arise instead from deviations of the burst  component from a blackbody.

We take the approach of assuming the burst component retains a consistent spectral shape, which we model as a blackbody plus power law. We returned to our "prototype" burst (see \S \ref{sec:OneBurst}) and fit the Eddington-limited and cooling tail spectra (stages 1, 2, and 3 in table \ref{tab:stage-defns}) with an absorbed blackbody plus powerlaw, with all parameters except $N_H$ variable. The mean power law index of these fits was 1.93$\pm$0.68. We then fit the spectra again, this time holding the power law index fixed at this value but allowing the normalization to still vary. As we are here assuming that the power law is intrinsic to the burst emission, rather than being a separately varying component, we must specify the power law normalization so that it contributes a fixed proportion of the total flux. Using a linear fit, we found that the normalization of the power law was best described by
\begin{equation}
\text{Power law norm}\approx \dfrac{K_{bb} \times kT^4}{3212},
\label{eqn:phenom-relation}
\end{equation}
where $kT$ is the blackbody temperature in keV and $K_{bb}$ is the normalization of the blackbody in km$^2$/100kpc$^2$. We fit the spectra a third time, this time tying the power law normalization to the blackbody parameters with the relation given by equation \ref{eqn:phenom-relation}, which fixes the shape of the burst emission.

We found the mean $\chi^2_\nu$ was 1.27$\pm$0.43 for the spectra of the Jun 15, 2001 burst from 4U~1636$-$536 using this fitting method, compared with 1.69$\pm$0.58 and 1.21$\pm$0.37 for the standard and variable persistent methods respectively. Thus, for this burst, the phenomenological model fits the data nearly as well as the variable persistent model. We then used the same model, with the same power law index and normalization relation, to fit every stage 1, 2, and 3 spectrum from all radius expansion bursts in our catalog. Restricting these to spectra where \XSPEC\ did not encounter fitting errors for any of the three methods, we found that our phenomenological model gave a mean $\chi^2_\nu$ of 1.43$\pm$1.10, compared with 1.45$\pm$0.82 and 1.21$\pm$0.60 for the standard and variable persistent methods respectively. The phenomenological model is thus globally no better than the standard approach. If there is a non-blackbody contribution to the burst component, it must differ between sources and change from burst to burst.  Furthermore, it would also have to resemble the persistent emission for every burst, at least superficially, or the variable $f_a$ method would not be able to consistently improve the reduced $\chi^2$; and this itself hints at a relationship with the persistent emission. Clearly any further investigation into such deviations must be physically motivated rather than phenomenological.

We then tested whether introducing a phenomenological change to the burst spectrum removes the need for a variable persistent emission. We fit the spectra of the Jun 15, 2001 burst from 4U~1636$-$536 using the variable persistent model, but replacing the blackbody burst component with the phenomenological spectrum. If our initial detection of a variable $f_a$ merely reflected the non-blackbody character of the burst component, then we would expect the detection to disappear. We still detect $f_a$ to vary significantly, though the values are consistently lower by roughly 20\% than for the original fits. Furthermore, the mean $\chi^2_\nu$ for these fits was 1.17$\pm$0.35, suggesting that allowing the persistent emission to vary improves the fits even when a non-blackbody model for the burst component is assumed. These results suggest very strongly that our detection of $f_a$ cannot be attributed to a confounding spectral effect intrinsic to the burst component.

\begin{figure}
\includegraphics[width=80mm]{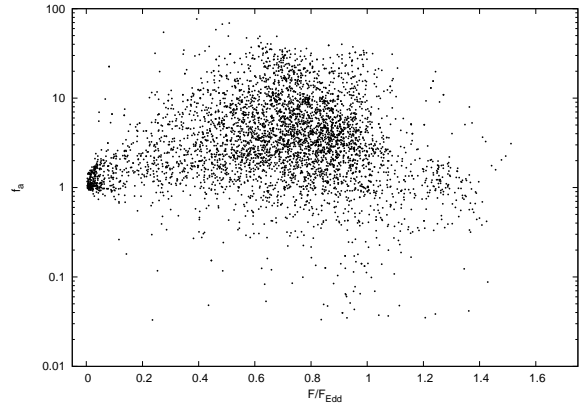}
\caption{$f_a$ against normalized burst flux for all Eddington-limited and cooling tail spectra. There is no obvious correlation between the two quantities.  There is a clearly visible clump at low flux and $f_a\approx 1$, showing that $f_a$ returns to its original level after the burst. There are also many points with high $f_a$ and normalized fluxes between  0.2 and unity, indicating that $f_a$ can remain elevated a considerable way down the cooling tail.}
\label{fig:favflux}
\end{figure}

If our variable $f_a$ fits are measuring a non-blackbody contribution to the burst spectrum then we should expect $f_a$ to increase with burst flux.  Suppose that a non-Planckian part of the burst component contributes a fixed fraction of the burst component flux and that our variable $f_a$ fits are trying to remove it. Then if the burst flux doubles, we would require $f_a$ to double also in order to fit out the non-Planckian part. We investigated this by comparing $f_a$ against $F/F_{Edd}$, where $F_{Edd}$ is the source's Eddington flux (see Appendix A) for all Eddington-limited and cooling tail spectra. It is clear from Figure \ref{fig:favflux} that there is no relationship between the quantities. To test whether plotting all the points together obscures a relationship present in individual bursts we selected  the Eddington-limited and cooling tail spectra from our prototype burst and three randomly selected other bursts, and performed Kendall rank correlation tests on their $f_a$ against their blackbody fluxes. No statistically significant correlation was present in any of these bursts. If the hypothesized non-blackbody contribution is present at the high energy tail, then we would expect an inverse correlation between blackbody temperature and $f_a$ because at low temperatures the blackbody drops out of the band detectable by $RXTE$, leaving only the hard tail to be fit. However, we find a slight \emph{positive} correlation of $\tau=0.031$ at $4.2\sigma$ significance using a Kendall rank correlation test.

Our  new fitting model holds the shape of the persistent emission fixed. Since the persistent emission of LMXBs changes shape according to whether it is in a high or low state \citep{HasingervanderKlis1989}, it is reasonable to think it may change shape temporarily if the accretion rate changes during a burst.  It is possible that spectral shape changes in the persistent spectrum contribute to the high $\chi^2$ in our approach. However, it is not clear that an accretion rate enhancement due to radiation drag will have the same effects on the persistent spectrum as the neutron star's usual movement around the Z-track, since they have very different physical causes. Indeed, \cite{intZandEtAl2013} found in their study of a burst from SAX~J1808.4$-$3658 that the persistent spectrum became harder, whereas ordinarily they would expect increased accretion to soften the spectrum.

\section{Comparison with theory}
\label{sec:WalkerComparison}

W92 performed 1D simulations of geometrically thin, axisymmetric irradiated accretion discs around a neutron star. Their models include the effects of viscosity and general relativity. W92's results predict that radiation torque from bursts can enhance the accretion rate by up to two orders of magnitude. If our $f_a$ is assumed to be entirely an accretion enhancement, then it corresponds identically to their quantity $\Delta\dot{M}_*/\dot{M}_*$ and is predicted to be inversely correlated with pre-burst accretion rate, accretion disc viscosity parameter $\beta$, neutron star spin frequency, and neutron star radius. All of W92's models assume a neutron star of mass $1.4M_\odot$.

\begin{figure}
\includegraphics[angle=270,width=80mm]{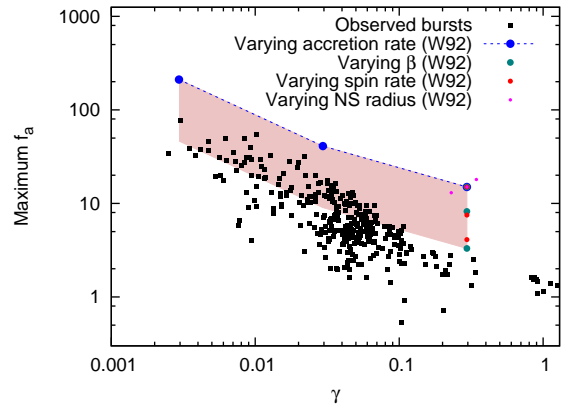}
\caption{Maximum $f_a$ against $\gamma$ (the pre-burst accretion rate as a fraction of \mdotedd) for all PRE bursts (black squares). Also plotted are the results of ten computer simulations performed by W92 (see their Table 1). The real bursts show generally lower $f_a$, but the same slope, as W92's nonrotating neutron star model. If a moderate $\beta$ and spin frequency of 300 to 600~Hz is assumed, then our results appear to agree with W92. The shaded region indicates the approximate area spanned by W92's models.  The bottom edge of the shaded box is uncertain due to the fact that none of W92's models vary both rotation rate and disc viscosity simultaneously, but it may extend further down than shown here.}
\label{fig:usvwalker}
\end{figure}

Table 1 of W92 lists peak accretion enhancements for the computed models; in Figure \ref{fig:usvwalker} we compare the maximum $f_a$  measured for each burst against $\gamma$, the pre-burst accretion flux as a fraction of the Eddington flux of the burst source. The Eddington flux for each neutron star is the mean peak flux of every PRE burst observed from that star; see Appendix A for the values and details of their calculation. Following G08 we measure $\gamma$ by integrating the chosen pre-burst persistent model between 2.5 and 25~keV and dividing the resulting flux by the burst source's Eddington flux. W92's Table 1 lists related quantities. W92's models begin with a non-rotating neutron star with radius 9km, $\gamma$ of roughly 0.3, and disc viscosity parameter, $\beta$, of $10^{-4}$. They then allow the spin frequency, accretion rate, $\beta$, and radius to vary in turn, while holding the other quantities fixed at their original values.

The disc viscosity parameter $\beta$ \citep{Coroniti1981} is similar to the Shakura-Sunyaev disc viscosity \citep{ShakuraSunyaev1973} but relates the viscosity to the gas pressure rather than the total pressure, which differs in discs which are radiation-pressure dominated. W92 gives accretion rates in dimensionless units: $\dot{m}=\dot{M}c^2/L_\text{Edd}$, where $\dot{M}$ is the mass accretion rate and $L_\text{Edd}$ is the Eddington luminosity, whereas we give accretion rates in terms of the energy release. Since the mass $M_*$ and radius $R_*$ of the neutron stars in W92 are specified, we have
\begin{equation}
\gamma=\dfrac{GM_*}{c^2R_*}\dot{m}.
\end{equation}

Our results in Figure \ref{fig:usvwalker} show a decrease in peak $f_a$ with increasing $\gamma$, and the slope is consistent with the predictions of W92's three points representing a nonrotating star with increasing accretion rate, while our peak $f_a$ values are significantly lower than those predicted for a nonrotating neutron star. If we assume a moderate $\beta$ and rotation frequency of about 300 to 600~Hz (e.g., \citealt{MunoEtAl2001}; see also Appendix A), then not only the observed correlation of peak $f_a$ with $\gamma$, but also the normalization, appear to be consistent with W92's predictions (see Figure \ref{fig:usvwalker}). The models predict that stellar rotation period and disc viscosity parameter $\beta$ have a large influence on peak $f_a$ but that neutron star radius apparently has little influence.  The upper edge of our measured points appears to be very roughly consistent with $\gamma f_{a}\lesssim 1$, implying that the accretion luminosity cannot greatly exceed $L_\text{Edd}$. This is consistent with the predictions of \cite{BurgerKatz1983} and \cite{MillerLamb1996}, who find that $\dot{M}_\text{Edd}$ is a natural cap on the accretion rate.  

\begin{figure}
\includegraphics[angle=270,width=80mm]{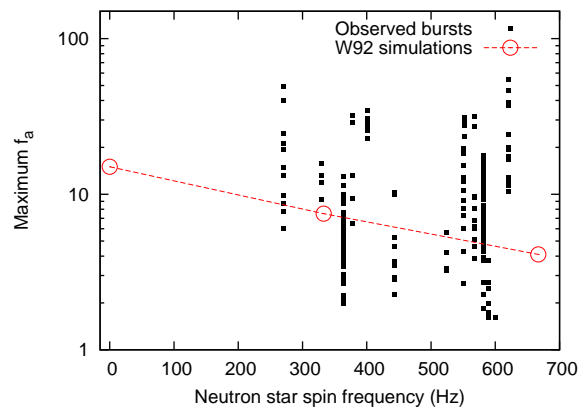}
\caption{Maximum $f_a$ against neutron star spin frequency for all PRE bursts arising from sources whose spin frequency is known (see Appendix A). Also plotted are the results of three computer simulations performed by W92 with spin frequencies of $10^{-5}$, 333, and 667 Hz. Although there is large scatter due to the dependence of $f_a$ on pre-burst $\dot{m}$ and $\beta$, our results are consistent in magnitude with W92's predictions. The anti-correlation of peak $f_a$ with spin period predicted by theory is not obviously visible in our data.}
\label{fig:maxfavspin}
\end{figure}

In Figure \ref{fig:maxfavspin} we show maximum $f_a$ against spin rate for all neutron stars in our sample whose spin rate is known. These are listed in Appendix A. W92's models predict a gradual decrease of maximum $f_a$ with spin frequency. Our data shows maximum $f_a$ values consistent in magnitude to W92's models, but it is difficult to discern any trend, because of the large scatter in individual sources (due to the additional dependence on $\gamma$ and the disc viscosity) and the small range of known neutron star spin frequencies (since these are all accreting neutron stars). We would require PRE bursts from slowly rotating neutron stars to better constrain this relationship.  A further difficulty is that W92's models do not reach the Eddington limit, so we are comparing PRE bursts in observations to non-PRE bursts in models.

 As the W92 models are one-dimensional, care must be taken in applying them to accreting neutron star systems in which three-dimensional effects are likely to be important. Further theoretical studies, preferably in three dimensions, would be very valuable.

\section{Discussion}
\label{sec:Discussion}

We have performed an observational investigation into whether or not the persistent emission contribution varies during photospheric radius expansion bursts. We allowed the pre-burst emission to vary, parametrized by a factor which we denote $f_a$. We detected a statistically significant increase in $f_a$ for nearly all the PRE bursts in the catalog, suggesting an enhanced (rather than suppressed) level of persistent emission during a burst. Our new method results in a significant improvement in the reduced $\chi^2$ of spectral fits compared to the standard model  but not to the level of formal statistical consistency for all the spectra.

Since the persistent emission is known to be an approximate measure of the accretion rate onto the neutron star, we interpret our results to indicate that the accretion rate onto the neutron star generally increases during bursts.  Obviously the persistent spectrum could also change shape in response to a varying accretion rate, and there are suggestions \citep{HomanEtAl2007} that the X-ray luminosity may not be proportional to the mass accretion rate. There are, however, currently no predictions about how the persistent emission spectrum may change in response to rapidly increasing radiation drag, so we make the simplest assumptions that it does not change shape and that $f_a\propto \dot{M}$. We have shown that peak $f_a$ measured during each burst anti-correlates with the pre-burst accretion rate with a slope consistent with that predicted by the theoretical models of \cite{Walker1992}, who investigated the effect of radiation drag on the accretion disc. If the effects of neutron star spin are accounted for, the magnitudes of the observed peak $f_a$ are also consistent with theory. This suggests that our detection of an increased pre-burst persistent emission reflects an enhanced accretion rate due to radiation torques on the accretion disc around the star.

All of W92's models are sub-Eddington, so some care must be taken in extrapolating their results to PRE bursts. However, W92 points out that the expanding atmosphere in a PRE burst should have very little angular momentum compared with the accretion disc. Radiation coupling between disc and envelope can thereby spin down the disc, increasing the accretion rate. This would also have the effect of spinning up the envelope and boosting its expansion. Such a mechanism may help explain why the atmosphere takes longer to return to the neutron star surface than to reach maximum radius. 

It may be argued that a deviation of the burst spectrum from a blackbody could mimic a variable accretion rate by introducing a high energy excess that our method then attempts to remove. If the deviation is of fixed size then any spurious $f_a$ measurement it causes will also anti-correlate with the pre-burst accretion rate $\gamma$, similarly to Figure \ref{fig:usvwalker}. However, we would then also expect $f_a$ to be constant with constant burst flux, but we found it to vary greatly during the Eddington-limited phase when the burst flux is approximately constant. We found no anti-correlation between $f_a$ and blackbody temperature, which would occur if the blackbody component drops out of \emph{RXTE}'s detection band and leaving only the deviation to be fit out. We also attempted to model a hard tail deviation in the burst spectrum using a power law with fixed index and normalization tied to the burst flux. This did not improve the spectral fits for the entire collection of bursts, and did not cause our detection of $f_a$ to be suppressed.  While in general we do not expect to see discrete spectral features in PRE bursts, in superexpansion bursts they can be visible \citep{intZandWeinberg2010}. We have found that in these events $f_a$ consistently drops rather than rises,  consistent with zero flux from the source. As suggested by \cite{intZandWeinberg2010} this may be due to the emission from the burst component dropping out of the band detectable by \emph{RXTE} together with the persistent emission being obscured by the superexpanding shell, though it is also possible that the vigorously expanding envelope disrupts the accretion disc and thereby temporarily halts accretion. An enhanced $f_a$ cannot be attributed to the presence of spectral features superimposed on the burst component continuum.  While we have attempted to exclude confounding spectral effects intrinsic to the burst component as an explanation for $f_a$ enhancements, we cannot rule out the possibility of other interpretations.

If our variable persistent fits causes the inferred photospheric radius at touchdown to differ systematically from the standard fits, then this would have implications for determinations of the equation of state of neutron star matter, such as those of \cite{OzelEtAl2009, GuverEtAl2010, SteinerEtAl2010}. We investigated this by taking the ratio of  the touchdown radii as determined by the variable $f_a$ and standard methods for all the PRE bursts in our sample and obtained a mean value of $0.97\pm0.11$. This indicates that the inferred touchdown radius can differ by $\sim$~10\%, and that there is a slight trend towards lower neutron star radii using the variable persistent method.  We have found that $f_a$ frequently remains elevated after the end of the Eddington-limited phase; this may have implications for studies that use the cooling tail to constrain the neutron star parameters (e.g., \citealt{GallowayLampe2012}).

\subsection{Structure of the accretion disc}

The details of the transfer of material from the inner edge of the accretion disc to the neutron star surface are still uncertain (e.g., \citealt{Bildsten1998}). It is not known if it occurs at a close inner boundary layer, or if there is some mechanism that regulates the infall. PRE bursts offer a means of settling this question: to be subject to radiation pressure the expanding photosphere must be optically thick and therefore must obscure our view of everything interior to it. The atmosphere expanding so far that it covers all of the emitting parts of the accretion disc would cause the persistent emission to be reprocessed inside the optically thick atmosphere and effectively become part of the burst emission. Even very modest radius expansion would hide the boundary layer due to this ''shrouding'', and cause $f_a$ to decrease, since the boundary layer is thought to be geometrically small and located at a radius $\sim 1.2R_*$ (e.g., \citealt{PophamSunyaev2001}). Thus our finding of consistently high $f_a$ during a burst argues against the existence of a thin boundary layer that remains near the surface of the star and dominates the persistent emission (see Figure \ref{figfavra}), but is consistent with a boundary layer that becomes much wider in response to increased luminosity as suggested by \cite{PophamSunyaev2001}.

\begin{figure}
\includegraphics[angle=270,width=80mm]{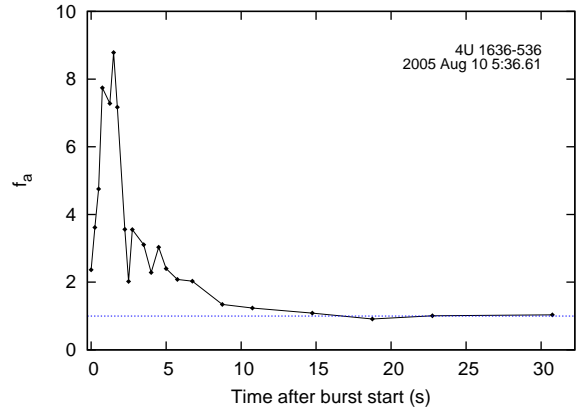}
\caption{A non-PRE burst from 4U~1636$-$536 observed on August 10 2005. The $f_a$ factor is significantly elevated, suggesting that the level of persistent flux rises for non-PRE bursts as well as PRE bursts. The time evolution of $f_a$ is similar to the one shown in \ref{figlayered}, indicating that the physical process is similar in both cases.}
\label{fig:nonPRE}
\end{figure}

We also note that performing the analysis method of this paper on non-PRE bursts may provide further information about the structure of the accretion disc. In these bursts the atmosphere does not expand and will not obscure any of the accretion disc. A careful comparison of non-PRE bursts against PRE bursts, which progressively obscure parts of the disc, might therefore give insights into structure and properties of the disc.  A preliminary investigation of a number of non-PRE bursts using the same analysis method indicates that the $f_a$ values increase during these events as well (see figure \ref{fig:nonPRE}), which would be expected if increased \mdot\ is caused by radiation drag as in W92's models, but which would not be expected if $f_a$ is tracking spectral changes caused by the emitting photosphere distending. A proper study of non-PRE bursts will be the subject of a subsequent paper.

\section*{Acknowledgments}

H.W. is supported by an APA postgraduate research scholarship. D.K.G. is the recipient of an Australian Research Council Future Fellowship (project FT0991598). This research utilizes preliminary analysis results from the Multi-INstrument Burst ARchive (MINBAR)\footnote{see \url{ burst.sci.monash.edu/minbar } }, which is supported under the Australian Academy of Science's Scientific Visits to Europe program, and the Australian Research Council's Discovery Projects and Future Fellowship funding schemes. This paper uses results provided by the ASM/RXTE teams at MIT and at the RXTE SOF and GOF at NASA's GSFC. This research has made use of data obtained through the High Energy Astrophysics Science Archive Research Center Online Service, provided by the NASA/Goddard Space Flight Center. We thank Tullio Bagnoli, Andrew Cumming, Alex Heger, Laurens Keek, Mark Walker, Michael Zamfir, and Jean in 't Zand for helpful conversations and suggestions. We are grateful to the anonymous referee, whose comments led to improvements in this paper.

\section{Appendix A}

Table \ref{tab:NHtable} shows the $N_H$ values we adopt for each burst source studied in this work, the reference from which we obtained that value, and the number of photospheric radius expansion bursts recorded by \emph{RXTE} from that source, including bursts we discarded due to source confusion or unsuitable spectral data (see \S \ref{sec:Pers-Models}).

We also give Eddington fluxes,  measured using the standard approach, for all sources that have PRE bursts recorded by either \emph{RXTE} (i.e. in the extended G08 catalog) or the Wide Field Camera on \emph{BeppoSAX}\footnote{see \url{ burst.sci.monash.edu/minbar }}.  The Eddington fluxes are the means of the peak fluxes for every PRE burst recorded for that source, weighted by the inverse square of the error of the measurement.

If only one PRE burst has been recorded for any source, we simply report the peak flux for that burst and the error on that measurement. For sources with two or more PRE bursts we calculate a reduced $\chi^2_\text{meas}$ assuming the source has constant Eddington flux. If $\chi^2_\text{meas}$ is consistent with a constant flux, we allow the flux to vary up and down such that it now has $\chi^2_\text{new}=1+\chi^2_\text{meas}$; the amount by which we vary it is the error on the original measurement. If $\chi^2_\text{meas}$ is \emph{not} consistent with constant Eddington flux we artificially scale the errors on the individual measurements until $\chi^2_\text{meas}=1$. Then we allow the flux to vary such that $\chi^2_\text{new}=2$.

The prolific burster 4U~1636$-$536 has PRE bursts that are known to be bimodal in flux \citep{EbisuzakiNakamura1988,GallowayEtAl2006}. Most of its PRE bursts are thought to take place in a pure helium atmosphere and have peak fluxes around ($68.6 \pm 5.5$)$\times$10$^{-9}\text{ erg cm}^{-2}\text{ s}^{-1}$. A few bursts have peak fluxes lower by a factor of $\approx$1.7, and these are believed to take place in a hydrogen-rich atmosphere (G08). We consider both regimes separately. The lower flux bursts include a tentatively identified PRE event recorded by \emph{RXTE} on 2000 Jan 22, 04:43:48 UT. We include this burst to calculate the mean Eddington flux for these low flux events, but exclude it for the rest of the analyses in this paper.

We also list neutron star spin frequencies, where known. These have been calculated either from burst oscillations (BO) (e.g., \citealt{Watts2012}), or from pulsar timing.

\begin{deluxetable*}{lrrlrccc}
\tabletypesize{\scriptsize}
\tablecaption{$N_H$ values chosen for spectral fits to burst data, Eddington fluxes, and spin periods
  \label{nHvalues} }
\tablewidth{0pt}
\tablehead{
\colhead{Source} & \colhead{$N_H$ ($10^{22}\,{\rm cm^{-2}}$)} & \colhead{No. of PRE bursts} & \colhead{Ref.} & \colhead{Eddington flux} & \colhead{Spin (Hz)} & \colhead{Method\tablenotemark{c}} & \colhead{Ref.}\\
 & & \colhead{\emph{RXTE} (\emph{WFC})\tablenotemark{d}} & \colhead{($N_H$)} & \colhead{ 10$^{-9}\text{ erg cm}^{-2}\text{ s}^{-1}$ } & & & \colhead{(Spin)}}
\startdata
  4U 0513$-$40     &   0.03 &  4 (3) &1 & $14.5\pm 3.5$\\
  EXO 0748$-$676   &   0.80 &  5 (0) &2 & $46.5\pm 4.6$&552&BO&27, 28\\
  2S 0918$-$549    &   0.35 &  2 (3) &3 & $119.2 \pm 12.4$\\
  4U 1608$-$522    &   0.89 & 17 (5) &4 & $ 172.2 \pm 21.8$&620&BO&27, 29, 30\\
  4U 1636$-$536 (high)\tablenotemark{a} &   0.25 & 76 (2) &5 & $73.9 \pm 6.8$ & 581 & BO & 27, 31, 32\\
  4U 1636$-$536 (low) & 0.25 &2\tablenotemark{b} (0) &5 & $41.5 \pm 1.2$ & 581 & BO & 27, 31, 32\\
  MXB 1658$-$298   &   0.20 & 12 (0) &6 & $17.0\pm 4.0$&567&BO&27, 33\\
  XTE 1701$-$462   &   2.00 &  2 (0) &7 & $43.4 \pm 1.4$\\
  4U 1702$-$429    &   1.87 &  5 (3) &8 & $87.7 \pm 4.5$&329&BO&27, 34\\
  4U 1705$-$44     &   1.90 &  4 (0) &9&$41.0 \pm 3.8$\\
  XTE J1710$-$281  &   0.40 &  3 (0) &10&$7.1 \pm 1.5$ \\
  4U 1722$-$30   &     0.78 &  3 (23) &1&$61.7\pm12.4$\\
  4U 1728$-$34     &   2.60 & 94 (0) &11&$95.0\pm 8.4$&363&BO&27, 35\\
  KS 1731$-$260    &   1.30 &  4 (3) &12&$48.6 \pm 5.6$&524&BO&27, 36, 37\\
  4U 1735$-$444    &   0.14 & 10 (0) &13&$34.2\pm5.6$\\
  GRS 1741.9$-$2853&  11.30 &  6 (1) &14& $35.3\pm 10.9$&589&BO&27, 38\\
  1A 1742$-$294    &   1.16 &  2 (0) &8& $37.8 \pm 1.4$\\
  SLX 1744$-$300    &  4.50 &  4 (3) &15 &$13.9\pm 3.1$\\
  GX 3+1           &   1.59 &  2 (0) &16&$60.0\pm1.4$\\
  SAX J1748.9$-$2021&  0.47 & 11 (0) &1&$38.0\pm6.0$&442&Pulsar&27, 39\\
  EXO 1745$-$248    &  3.80 &  2 (0) &1&$69.0\pm2.8$\\
  4U 1746$-$37      &  0.26 &  3 (0) &1&$5.4\pm0.7$\\
  SAX J1747.0$-$2853 & 8.80 & 11 (2) &17&$52.5\pm7.1$\\
  IGR J17473$-$2721 &  3.80 &  3 (0) &18&$113.5\pm12.1$\\
  IGR J17498$-$2921 &  1.20 &  1 (0) &8&$51.6\pm1.6$&401&Pulsar&27, 40\\
  XTE J1759$-$220   &  2.84 &  3 (0) &8&$15.7\pm0.8$\\
 SAX J1750.8$-$2900 &  0.90 &  2 (1) &8&$54.1\pm2.1$&601&BO&27, 30, 41\\
 GRS 1747$-$312     &  1.39 &  3 (0) &1&$13.4\pm4.4$\\
 SAX J1808.4$-$3658 &  0.12 &  8 (2) &19&$230.1\pm13.2$&401&Pulsar&27, 42\\ 
 XTE J1810$-$189    &  4.20 &  1 (0)  &20&$54.2\pm1.8$\\
 SAX J1810.8$-$2609 &  0.35 &  1 (1) &21&$111.2\pm3.0$\\
   GX 17+2          &  1.90 &  2 (0) &22&$15.5\pm0.5$\\
  4U 1820$-$303     &  0.16 & 16 (34) &1&$60.5\pm4.0$ \\
  XB 1832$-$330     &  0.05 &  1 (0) &1&$33.7\pm4.4$\\
  HETE J1900.1$-$2455& 0.16 &  7 (0) &23&$123.9\pm10.6$&378&Pulsar&27, 43\\
  Aql X$-$1          & 0.40 & 14 (0) &24&$99.6\pm21.3$&550&Pulsar&27, 44\\
  XB 1916$-$053     &  0.32 & 12 (0) &25&$30.6\pm3.6$&270&BO&27, 45\\
  4U 2129+12        &  0.03 &  1 (1) &1&$40.8\pm1.6$\\
  Cyg X$-$2         & 0.05  &  8 (0) &26&$13.1\pm2.1$\\
  Ser X$-$1          &  0.38 & 6 (0) &26&$29.4\pm7.1$\\
\enddata
\tablenotetext{a}{PRE burst fluxes from 4U~1636$-$536 are bimodal, separated by a factor of $\approx$1.7, thought to be due to bursts occurring in different atmospheric compositions (\citealt{EbisuzakiNakamura1988,GallowayEtAl2006}). Mean Eddington fluxes for both regimes are given.}
\tablenotetext{b}{This includes a tentatively identified PRE burst that is not included in the analysis in the rest of the paper.}
\tablenotetext{c}{BO = Burst Oscillation, Pulsar = Pulsar Timing}
\tablenotetext{d}{WFC = \emph{BeppoSAX} Wide Field Camera}
\tablerefs{1. \cite{KuulkersEtAl2003}, 2. \cite{HomanEtAl2003}, 3. \cite{JuettEtAl2001}, 4. \cite{KeekEtAl2008}, 5.\cite{AsaiEtAl2000}, 6. \cite{OosterbroekEtAl2001}, 7. \cite{LinEtAl2009}, 8. J. in't Zand, private communication (2012), 9. \cite{PirainoEtAl2007}, 10. \cite{YounesEtAl2009}, 11. \cite{DaiEtAl2006}, 12. \cite{CackettEtAl2006}, 13. \cite{AugusteijnEtAl1998}, 14. \cite{SakanoEtAl2002}, 15. \cite{MoriEtAl2005}, 16. \cite{OosterbroekEtAl2001b}, 17. \cite{WernerEtAl2004}, 18. \cite{AltamiranoEtAl2008}, 19. \cite{WangEtAl2001}, 20. \cite{KrimmEtAl2008}, 21. \cite{NatalucciEtAl2000}, 22. \cite{FarinelliEtAl2007}, 23. \cite{Campana2005}, 24. \cite{CampanaStella2003}, 25. \cite{ChurchEtAl1998}, 26. \cite{AsaiEtAl2000}, 27. \cite{Watts2012}, 28. \cite{GallowayEtAl2010b}, 29. \cite{HartmanEtAl2003}, 30. G08, 31. \cite{StrohmayerEtAl1998}, 32. \cite{StrohmayerMarkwardt2002}, 33. \cite{WijnandsEtAl2001}, 34. \cite{MarkwardtEtAl1999}, 35. \cite{StrohmayerEtAl1996}, 36. \cite{SmithEtAl1997}, 37. \cite{MunoEtAl2000}, 38. \cite{StrohmayerEtAl1997}, 39. \cite{AltamiranoEtAl2008}, 40. \cite{PapittoEtAl2011}, 41. \cite{KaaretEtAl2002}, 42. \cite{ChakrabartyEtAl2003}, 43. \cite{WattsEtAl2009}, 44. \cite{ZhangEtAl1998}, 45. \cite{GallowayEtAl2001}}
\label{tab:NHtable}
\end{deluxetable*}

\bibliography{/home/hworpel/PhD/phdbib.bib}
\end{document}